\documentclass[aps,prb,onecolumn,showpacs,longbibliography,final,notitlepage,floatfix,groupedaddress]{revtex4-1}
\usepackage{amssymb}
\usepackage{amsmath}
\usepackage{color}
\usepackage{hyperref}
\usepackage{graphicx}
\usepackage{subfigure}

\makeatletter

\newcommand\la{\langle}
\newcommand\ra{\rangle}

\def\ket#1{{| #1 \rangle}}
\def\bra#1{{\langle #1 |}}
\newcommand{\braket}[2]{\langle #1 |#2\rangle}

\def\ii{{\rm i}}

\def\1{\mathbb{I}}

\def\l1{\lambda_1}

\def\lp{\lambda_{\mathrm{ps}}}

\begin{document}

\title{Boom and bust cycles due to pseudospectra of matrices with unimodular spectra}

\author{Junaid Majeed Bhat, Ja\v{s} Bensa, and Marko \v Znidari\v c}

\affiliation{Department of Physics, Faculty of Mathematics and Physics, University of Ljubljana, 1000 Ljubljana, Slovenia}

\date{\today}

\begin{abstract}
We discuss dynamics obtained by increasing powers of non-normal matrices that are roots of the identity, and therefore have all eigenvalues on the unit circle. Naively, one would expect that the expectation value of such powers cannot grow as one increases the power. We demonstrate that, rather counterintuitively, a completely opposite behavior is possible. In the limit of infinitely large matrices one can have an exponential growth. For finite matrices this exponential growth is a part of repeating cycles of exponential growths followed by exponential decays. The effect can occur if the spectrum is different than the pseudospectrum, with the exponential growth rate being given by the pseudospectrum. We show that this effect appears in a class of transfer matrices appearing in studies of two-dimensional non-interacting systems, for a matrix describing the Ehrenfest urn, as well as in previously observed purity dynamics in a staircase random circuit.
\end{abstract}


\maketitle

\section{Introduction}

Matrices among others describe the most general linear relation between two finite-dimensional vectors and are as such of utmost importance in physics. Often the relevant quantity is expressed in terms of powers of matrix $A$, for instance as
\begin{equation}
  f(t) = \la w | A^t | v \ra,
  \label{eq:f}
\end{equation}
where $A$ is a $N \times N$ matrix and $\bra{w}$ and $\ket{v}$ are two vectors. We would expect that the behavior of $f(t)$ at large times $t$ will be dominated by the largest eigenvalue of $A$ (suppose $A$ is diagonalizable), that is $f(t \gg 1) \sim |\lambda_1|^t$, where $\lambda_1$ is the largest eigenvalue in modulus. This intuition is based on our experience with normal matrices ($[A,A^\dagger]=0$), for instance Hermitian, unitary, etc.. Namely, such matrices can be diagonalized by a unitary transformation, and therefore in an appropriate orthogonal basis the dynamics of $A^t$ is given by the powers of eigenvalues. Specifically, if all eigenvalues are unimodular, that is $|\lambda_j|=1$, which will be the case for matrices discussed in this paper, the above reasoning would suggest that $f(t)$  cannot grow asymptotically and can be at most of order ${\cal O}(1)$.

\begin{figure}[h!]
        \begin{center}
        \includegraphics[width=0.4\textwidth,page=4]{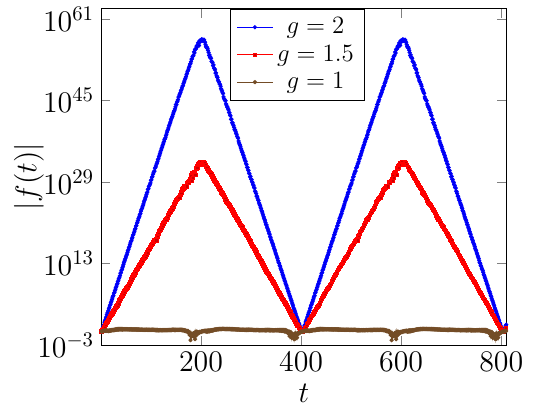} 
        \caption{Periodic exponential growth and exponential decay of $|f(t)|$, Eq.~\ref{eq:f}. For $g\neq1$, we see a periodic behavior with exponential growth followed by an exponential decay within each period. $A$ is a $400\times 400$ matrix given by Eq.~\ref{eq:T}. All eigenvalues of $A$ are unimodular, $|\lambda_j|=1$, while the largest value in the pseudospectrum is $\lp=g$, see Sec.~\ref{sec:simple} for more details. $\bra{w}$ and $\ket{v}$ are normalized random vectors with real components.}
    \label{fig:example}
    \end{center}
\end{figure}

We shall show that this intuition can be completely misleading for non-normal matrices. We find a class of matrices where $f(t) \sim {\cal O}(1)$ is violated in a maximal manner: despite all eigenvalues being unimodular, in the limit $N \to \infty$ not only does $f(t)$ grow with $t$, it grows so exponentially and can therefore become huge for large $N$. We will treat matrices $A$ that are roots of the identity matrix, $A^T=\1$, meaning that all eigenvalues are on a unit circle, $\lambda_j={\rm e}^{\ii \varphi_j}$, and, furthermore, that all phases are commensurate, $T \varphi_j = 0 \pmod{2\pi}$. An important consequence of the root property is that $f(t)$ is periodic, $f(t+T)=f(t)$. Nevertheless, $f(t)$ will exhibit surprising alternating cycles of exponential growth (boom), immediately followed by an exponential decay (bust). An example is shown\footnote{We plot the absolute value of $f(t)$ because it is negative at few points around the maximum.} in Fig.~\ref{fig:example}. 
The length of the first boom cycle for this specific matrix is given by $T/2=N/2$ and therefore goes to infinity as $N \to \infty$. In the limit when one first lets $N \to \infty$ only the exponential growth phase is observed, seemingly at complete odds with an ``intuitive'', but incorrect, $f(t) \sim {\cal O}(1)$. How can that be?

The resolution lies in the behavior of matrix norms for non-normal matrices. Let us denote by $\rho$ the spectral radius of $A$, $\rho={\rm max}_j(|\lambda_j|)$. For normalized $w$ and $v$ we can rigorously bound $f(t) \le \lVert A^t \rVert$, however, it is in general incorrect to argue that $\lVert A^t \rVert \sim \rho^t$ for large $t$, even-though the exact Gelfand's formula~\cite{Horn} does say that $\lim_{t \to \infty} \lVert A^t \rVert^{1/t}=\rho$. The Gelfand's formula is about the limit when one at a fixed $N$ first takes $t \to \infty$, and in principle does not tell us anything about the other order of limit when one first takes $N \to \infty$. The bounds one can make~\cite{trefethen} at a finite $t$ are $\rho^t \le \lVert A^t \rVert \le \lVert A \rVert^t$, or~\footnote{Using the spectral norm~\cite{Horn} one could furthermore bound $\lVert A \rVert_2 \le s_{\rm max}(A)$, where $s_{\rm max}(A)$ is the largest singular value of $A$.}, alternatively for diagonalizable $A$, $V^{-1}AV=\Lambda$ with diagonal $\Lambda$. Another exact inequality is $\lVert A^t \rVert \le \kappa \rho^t$, where $\kappa=\lVert V \rVert \lVert V^{-1} \rVert$ is the condition number of $V$. If we take the spectral norm of $V$, we have $\kappa=s_{\rm max}(V)/s_{\min}(V)$, where $s(V)$ are singular values of $V$. Thus, we see that if the matrix $A$ is normal we have $\kappa=1$ (because $V$ is unitary), whereas for general matrix $\kappa$ can be arbitrarily large which can render the inequality $f(t) \le \kappa \rho^t$ rather useless at finite $t$. Looking again at Fig.~\ref{fig:example} we can now understand how the surprising behavior $f(t)$ arises: in the limit $\lim_{N \to \infty}\lim_{t \to \infty}$, the long time average of $f(t)$ is constant, which is in accordance with $\rho=1$. However, in the limit $\lim_{t \to \infty}\lim_{N \to \infty}$ one instead observes an exponential growth that lasts until infinite times, ``allowed'' by the condition number $\kappa$ that diverges exponentially with $N$.

The question though remains what determines this exponential growth rate; it is clearly not the eigenvalues which are all $|\lambda_j|=1$? It is well known that the spectrum of non-normal matrices can be badly conditioned, i.e., very sensitive to perturbations. In such situations one should instead look at the pseudospectrum~\cite{trefethen}, which on the other hand is stable. This is also the case for our matrices; we will see that the initial growth rate of $f(t)$ is given by the largest pseudoeigenvalue $\lp$ in the pseudospectrum of $A$.

The pseudospectrum can be defined in various equivalent ways, for instance, as points $z$ in the complex plane where $(A-z\1)$ is almost singular, or as a spectrum of a slightly perturbed matrix (which has a well defined nontrivial limit when perturbation strength goes to zero), see Ref.~\onlinecite{trefethen} for more details. For normal matrices, the spectrum and the pseudospectrum coincide, whereas for non-normal ones they can be different. If the spectrum and pseudospectrum are different this means that (at least some) eigenvalues are very sensitive to perturbations. Because the sensitivity of a simple eigenvalue can be bounded by the eigenvalue condition number~\cite{Horn} $\kappa_j=\lVert R_j \rVert \lVert L_j \rVert/\langle L_j | R_j \rangle$ as $|\lambda_j(\epsilon)-\lambda_j| \le \kappa_j \epsilon$, where $\epsilon$ is the norm of the perturbation, and $R_j$ and $L_j$ are right and left eigenvectors, for such non-normal matrices the condition number diverges (when $N \to \infty$). There have been a number of previous observations that the pseudospectrum can be important in physics. One such case is that of a hydrodynamic instability and the notoriously difficult to understand critical Reynolds number at which the transition to turbulence happens. Using a linear stability analysis of a non-linear Navier-Stokes equation one finds~\cite{tref93,Boberg88} that the linearized operator is non-normal with highly unstable eigenvalues which can lead to a much earlier transition to turbulence than suggested by its eigenvalues. Pseudospectra have been shown to be one way to understand~\cite{okuma20} the non-Hermitian skin effect and in particular quasi-zero modes. Long-lived boundary edge modes in Lindblad master equations can be also related to non-normality and pseudospectra~\cite{viola21,viola23}. Also in Lindblad master equations it can happen that the relaxation rate is not given by the spectral gap and/or is sensitive to boundary conditions~\cite{pre15,song19,mori20,ueda21,mori21,mori23,clerk23}; such effects, although not discussed as such, can be likely traced to a diverging condition number of a non-Hermitian Liouvillian resulting in differing spectra and pseudospectra. A two-step relaxation in which the initial relaxation phase is surprisingly not given by the largest eigenvalue, and which was first observed numerically in purity and out-of-time-ordered correlation function dynamics~\cite{PRX,PRA,PRR,floq23}, can be also explained in terms of the pseudospectrum~\cite{marko22,marko23,preprint} (an alternative elegant explanation is possible using a membrane picture~\cite{zhou23}). Pseudospectra have been used also in non-Hermitian ${\cal PT}$-symmetric quantum mechanics~\cite{krejcirik15}, and is the quantity that determines a transient behavior, for instance in non-Hermitian photonics~\cite{makris21}. Considering that the pseudospectrum (a diverging condition number) is important for the sensitivity of eigenmodes, it has been used to discuss black hole stability~\cite{sheikh21}, and in a suggestion on how to define stable phases of open systems~\cite{sarang23}. For uses of pseudospectra outside of physics see also Ref.~\onlinecite{trefethen}.

Finally, let us briefly comment on the type of non-normal matrices that we study and whether they are relevant for physics. While normal matrices are obviously important, we note that often one has to deal with non-unitary and non-Hermitian matrices, even though the whole evolution is unitary. This happens, for instance, when the system is open (coupled to the environment), or if we do a coarse-graining or trace out some degrees of freedom. In a many-body system this might in fact typically be the case of interest because the full exponential complexity of unitary evolution is unobservable. In such context taking first the limit $N \to \infty$ (akin to the thermodynamic limit) is also the relevant order of limits. A simple example of a matrix that falls into our class of non-normal roots of identity is a tilted Pauli $\sigma^{\rm x}$ matrix, that is $X=\left( \begin{smallmatrix} 0 & g \\ 1/g & 0\end{smallmatrix} \right)$. One can easily check that $X^2=\1$, eigenvalues are $\lambda_j=\{-1,1\}$ and therefore $\rho=1$. Though, $X$ is not normal. One iteration can increase the norm of a vector by a factor $g$, e.g. $X\{0,1\}=g\{1,0\}$, being due to $s_{\rm max}(X)=g$. However, the matrix is of size $2$ and no really interesting effects can be observed in $f(t)$; in order to observe the exponential growth large $N$ is required.

In the next Section~\ref{sec:simple} we present one such example with motivation coming from transfer matrices and transport. We will be able to compute $f(t)$ exactly for a simple specific choice of $\bra{w}$ and $\ket{v}$, showing all the interesting features obtained for generic $w$ and $v$ like in Fig.~\ref{fig:example}. In Section \ref{sec:other} we present another mathematical example that manifests a similar periodic behavior, as well as briefly discuss purity evolution for a specific random quantum circuit and find some similarities between a strange behavior of purity observed in Ref.~\onlinecite{PRX} and $f(t)$ studied here.

\section{Exact solution of a simple example}
\label{sec:simple}

In this Section, we consider the matrix for which $f(t)$ exhibits periodic booms and busts shown in Fig.~\ref{fig:example}. Our simple example allows the computation of $f(t)$ analytically for specific $\bra{w}$ and $\ket{v}$ and thereby demonstrates that non-normal matrices that satisfy $A^T=\1$ can give rise to the mentioned counterintuitive dynamics. The matrix $A$ of size $N\times N$ has a $2\times 2$ block structure,
	\begin{equation}
	A=\begin{pmatrix}
	B && \1 \\-\1 &&0
	\end{pmatrix},
  \label{eq:T}
	\end{equation}
where $\1$ is the $M \times M$ identity matrix, $M=N/2$, and $B$ is an $M\times M$ tridiagonal Toeplitz matrix of the form
	
	\begin{equation}
	B=\begin{pmatrix}
	0  & 1/g&0  &0& \dots\dots& 0&0 \\
	g & 0 & 1/g&0 & \dots\dots&0&0 \\
	0 & g& 0 & 1/g &\dots \dots &0&0\\
	\vdots&\vdots &\vdots & \ddots&&\vdots&\vdots \\
	\vdots&\vdots&\vdots& & \ddots &\vdots&\vdots \\
	\vdots&\vdots&\vdots&\dots& g&0 & 1/g\\
	0&0 & 0&\dots&\dots&g & 0 
	\end{pmatrix}.
 \label{eq:B}
	\end{equation}
Matrices as those in Eq.~\ref{eq:T} occur while inverting block tridiagonal matrices~\cite{molinari1997transfer,reuter2012efficient},  and therefore are important in certain physical systems. For example,  in nearest-neighbor tight-binding lattice systems such matrices relate the components of the wavefunction on three consecutive hyperplanes of the lattice~\cite{lee1981simple}. To understand this further, consider a two-dimensional nearest-neighbor tight-binding Hamiltonian defined on a rectangular lattice of size $L\times W$ with open boundary conditions on its edges, and let $\Psi(x,y)$ be the wavefunction amplitude of the eigenstate of energy $E$ at the lattice site $(x,y)$. Since the hoppings are only between nearest neighbors, the Schrodinger equation for the eigenstate $\Psi(x,y)$ away from the edges at $x=1$ and $x=L$ can be written as,
\begin{equation}
    H_{i,i+1}^\dagger \psi(i-1)+H_{i,i}\psi(i)+H_{i,i+1}\psi(i+1)=E\psi(i)
    \label{eq:A_example}
\end{equation}
where $\psi(i)$ is a column vector containing the wavefunction components at $x=i$, i.e.  $\psi(i)=(\Psi(i,1),\Psi(i,2),...,\Psi(i,W))^T$, and $H_{i,j}$ is an $W\times W$ matrix that describes the hopping from the lattice sites at $x=i$ to the lattice sites at $x=j$.  Eq.~\ref{eq:A_example} can be written as,
\begin{equation}
    \begin{pmatrix}
        \psi(i+1)\\
        \psi(i)
    \end{pmatrix}=\begin{pmatrix} H_{i,i+1}^{-1}(E-H_{i,i}) && -H_{i,i+1}^{-1} H_{i,i+1}^\dagger  \\\1 && 0
        
    \end{pmatrix}\begin{pmatrix}
        \psi(i)\\
        \psi(i-1)
    \end{pmatrix}.\label{eq:tbeq1}
\end{equation}
 For the case of uniform hoppings, $n^{th}$ power of the matrix in Eq.~\ref{eq:tbeq1} relates the wave function components at $x=i$ and $x=i-1$ to the wavefunction components at $x=i+n$. Moreover, the powers of this matrix can also be used to determine the matrix elements of the single-particle Green's function for this system~\cite{reuter2011probing} which then determines two-point correlation functions, and transport properties such as conductance, current density, etc~\cite{dhar2006nonequilibrium}. We see that for $H_{i,i+1}^\dagger=H_{i,i+1}$, the matrix in the above equation reduces to the same form as $A$ from Eq.~\ref{eq:T} with $H_{i,i+1}^{-1}(E-H_{i,i})$ in place of $B$. It is possible to make simple choices of matrices $H_{i,i+1}$ and $H_{i,i}$ for $E=0$ such that our $B = H_{i,i+1}^{-1}(E-H_{i,i})$.

\begin{figure}[t!]
        \begin{center}
        \includegraphics[width=0.4\textwidth,page=2]{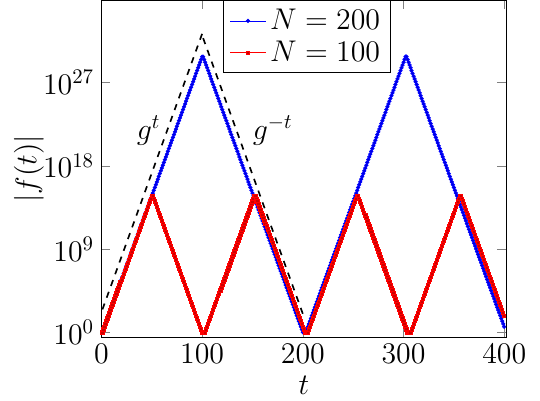} 
        \caption{Time evolution of $f(t)$ (Eq.~\ref{eq:f}) for special choices of the vectors $\ket{v}$ and $\bra{w}$ with components $w_p=1$  and $v_p=\delta_{1,p}$, $p = 1,2,\dots,N$. The plot shows results for $g=2$ and two different matrix sizes $N=100$ and $N=200$. The behavior in this figure is similar to the behavior from Fig.~\ref{fig:example} for random $\bra{w}$ and $\ket{v}$. The exponential growth is given by $g^t$, where $g$ is the maximum value in the pseudospectrum of $A$. The exponential decay is instead given by $(1/g)^t$.}
    \label{fig:growth}
    \end{center}
\end{figure}

Our goal is to study the behavior of the variable $f(t)=\bra{w}A^t\ket{v }$ with $t$. As seen in Fig.~\ref{fig:example}, when $g\neq 1$ any random choice of $\ket{v}$ and $\bra{w}$ gives an exponential growth in $t$, subsequently followed by an exponential decay. To understand this behavior analytically, we make a particular choice of the vectors $\ket{v}$ and $\bra{w}$ that retain the features observed in Fig.~\ref{fig:example} and also allow evaluation of $f(t)$ analytically. A simple (not normalized) choice of such vectors is given by $w_p=1$ and $v_p=\delta_{1,p}$, $p=1,2,\dots,N$. Fig.~\ref{fig:growth} shows a numerical computation of $f(t)$ with this choice of $\ket{v}$ and $\bra{w}$.  We see essentially the same periodic behavior as observed in Fig.~\ref{fig:example}, first exponential growth $g^t$, followed by an exponential decay $\sim g^{-t}$. We shall calculate $f(t)$ analytically by using the spectral decomposition of $A$, simplifying the resulting sums, as well as show that the $g$ in the exponential growth is exactly equal to the largest eigenvalue in the pseudospectrum of $A$.

Since $B$ is a simple tridiagonal Toeplitz matrix, its  left and right eigenvectors corresponding to the eigenvalue $\mu_k$ are readily calculated\cite{toeplitz}
\begin{align}
              \mu_k&=2\cos \frac{k\pi}{M+1},\\
 	      r_j^{(k)}&=g^{j}\sin \left( \frac{k\pi j}{M+1}  \right),\\
 	      l_j^{(k)}&=\frac{2}{M+1}g^{-j}\sin \left( \frac{k\pi j}{M+1} \right),\qquad k=1,2,\dots,M,
    \end{align}
where $k$ is an eigenvalue and eigenvector index, while $j=1,2, \dots,M$ runs through the components. Note that left and right eigenvectors are exponentially localized at opposite edges of the system, and are normalized such that $\langle l^{(k)} | r^{(k')}\rangle=\delta_{kk'}$.
\begin{figure}[h] 
    \centering
        \subfigure{\includegraphics[width=0.7\textwidth,page=3]{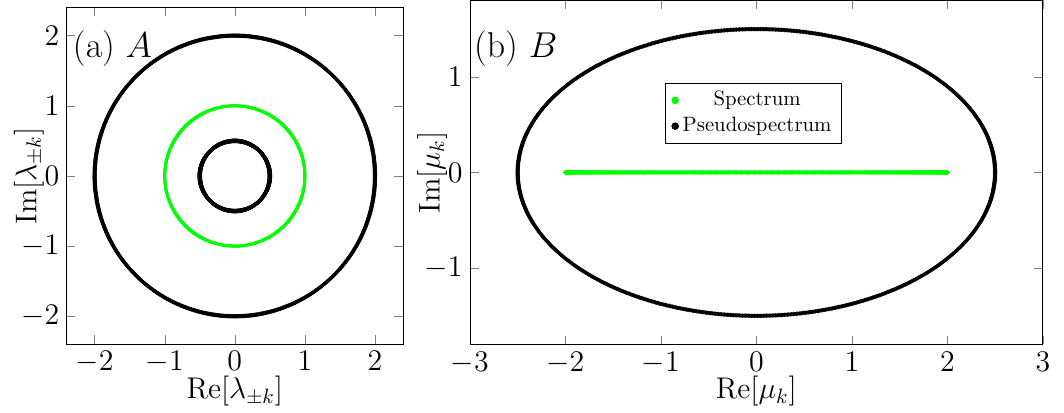}}
        \caption{The spectrum and pseudospectrum of the matrices $A$ (frame (a)) and $B$ (frame (b)), with $g=2$ and in the limit of infinite matrix size.}
    \label{fig:spect}    
\end{figure}

Now let us compute eigenvectors and eigenvalues of $A$. Let us denote by $\mathbf{x}=(\mathbf{x}_1,\mathbf{x}_2)^\mathbf{T}$ a right eigenvector of $A$ with eigenvalue $\lambda$. We then rewrite the equation $A\mathbf{x}=\lambda\mathbf{x}$ as the following two  equations
    \begin{align}
         B\mathbf{x}_1+\mathbf{x}_2&=\lambda\mathbf{x}_1\label{eq:eval1},\\
          -\mathbf{x}_1&=\lambda\mathbf{x}_2\label{eq:eval2}.
    \end{align}
Eq.~\ref{eq:eval2} gives $\mathbf{x}_2=-\frac{1}{\lambda}\mathbf{x}_1$, and plugging this into Eq.~\ref{eq:eval1} results in $B \mathbf{x}_1=(\lambda+\frac{1}{\lambda})\mathbf{x}_1$. Therefore, $\mathbf{x}_1$ is a right eigenvector of $B$ with eigenvalue $\mu=\lambda+(1/\lambda)\implies \lambda^2-\mu\lambda+1=0$. We see that the eigenvalues of $A$ are related to the eigenvalues of $B$ via a quadratic equation, and therefore each eigenvalue of $B$ will give two eigenvalues of $A$. Let $\mathbf{x}_1=r^{(k)}$ so that $\mu=\mu_k$, and let $\lambda_{\pm k}$ be solutions of the quadratic equation with $\mu=\mu_k$, then the eigenvalues of $A$ are 
    \begin{equation}
        \lambda_{\pm k}= \frac{\mu_k\pm i \sqrt{4-\mu_k^2}}{2}=e^{\pm i\frac{k\pi}{M+1}}\label{eq:specrelation}.
    \end{equation}
The corresponding right eigenvectors $R^{(\pm k)}$ are given by 
\begin{equation}
R^{(\pm k)}=\begin{pmatrix}
 	      r^{(k)}\\ -\lambda_{\mp k} r^{(k)}
 	      \end{pmatrix}.
\end{equation}
From Eq.~\ref{eq:specrelation} we see that all eigenvalues of $A$ lie on the unit circle, and that furthermore they are a $(2M+2)^{th}$ root of unity. Therefore, one has $A^{T}=\1$ with
\begin{equation}
  T=N+2.
  \label{eq:TT}
\end{equation}
A similar calculation determines the left eigenvectors
  \begin{align}
  L^{(\pm k)}=\frac{1}{1-\lambda_\pm^{-2}}\begin{pmatrix}
 	      l^{(k)}\\ \lambda_{\mp k} l^{(k)}
 	      \end{pmatrix}.
    \end{align}
The normalization is such that $\bra{L^{(\pm k)}}$ and $\ket{R^{(\pm k)}}$ form a biorthonormal system. 

The pseudospectrum of a Toeplitz operator, $C_{i,j} = c_{i-j}$, i.e. of an infinite-size matrix, is given by $\sum_{p=-\infty}^{\infty} c_p \mathrm{e}^{\ii p \phi}$, where $\phi \in [ 0,2\pi ]$ \cite{trefethen}. In our case, $B$ is a tridiagonal Toeplitz matrix with $c_{-1} = g$, $c_0= 0$ and $c_1 = 1/g$, therefore if $B$ is of infinite size, its pseudospectrum is $g \mathrm{e}^{-\ii \phi}+ (1/g) \mathrm{e}^{\ii \phi}$. The latter expression defines an ellipse with semi-major and semi-minor axes of length $g\pm 1/g$, respectively. Based on numerical evidence (not shown), we believe that the pseudospectrum of $A$ is related to the pseudospectrum of $B$ via the same relation as in Eq.~\ref{eq:specrelation}. This relation now maps the ellipse into two concentric circles given by $g e^{i\phi}$ and $(1/g)e^{-i\phi}$, $\phi \in [ 0,2\pi ]$ which constitute the pseudospectrum of $A$. We show the plots of the spectrum and the pseudospectrum of the two matrices in the limit $N \rightarrow \infty$ in Fig.~\ref{fig:spect}. The largest pseudoeigenvalue in the pseudospectrum of $A$ is therefore
\begin{equation}
  \lp=g.
  \label{eq:lps}
\end{equation}

Now that we know the left and right eigenvectors of $A$, we are ready to construct an explicit expression of $f(t)$ for our special choice of $\ket{v}$ and $\bra{w}$ using the spectral decomposition of $A$. A similar spectral calculation has been performed in Ref.~\onlinecite{marko23} for a more complicated Toeplitz matrix, and in Ref.~\onlinecite{preprint} for a tridiagonal one. We get
    \begin{align}
 	  f(t)&=\sum_{k=1}^M \mathrm{e}^{i\frac{k\pi}{M+1} t} \braket{w}{R^{(+ k)}}\braket{L^{(+ k)}}{v}
        +\sum_{k=1}^M \mathrm{e}^{-i\frac{k\pi}{M+1} t} \braket{w}{R^{(- k)}}\braket{L^{(- k)}}{v}\\\label{eq:sum1}
        &\notag= \sum_{k=1}^M \mathrm{e}^{i\frac{k\pi}{M+1} t} (1-\mathrm{e}^{-i \frac{k\pi}{M+1}})\sum_{l=1}^M g^{l}\sin\frac{k\pi l}{M+1} \frac{1}{1-\mathrm{e}^{-i\frac{2\pi k}{M+1}}}\frac{2}{1+M} g^{-1} \sin \frac{k\pi}{M+1}\\&+ \sum_{k=1}^M \mathrm{e}^{-i\frac{k\pi}{M+1} t} (1-\mathrm{e}^{i \frac{k\pi}{M+1}})\sum_{l=1}^M g^{l}\sin\frac{k\pi l}{M+1} \frac{1}{1-\mathrm{e}^{-i\frac{2\pi k}{M+1}}}\frac{2}{1+M} g^{-1} \sin \frac{k\pi}{M+1},
    \end{align}
where in Eq.~\ref{eq:sum1} we have put in the forms of the vectors $\bra{R^{(\pm k)}}$, $\ket{L^{\pm k}}$, $\bra{w}$ and $\ket{v}$ explicitly. The two sums over $k$ can be united in a single sum where $k$ runs from $-M$ to $M$. By doing so, we implicitly included a term for $k=0$ which evaluates to zero. 
 Therefore, Eq.~\ref{eq:sum1} can be written as
\begin{equation}
    f(t) = \sum_{k=-M}^M \mathrm{e}^{i\frac{k\pi}{M+1} t} (1-\mathrm{e}^{-i \frac{k\pi}{M+1}})\sum_{l=1}^M g^{l}\sin\frac{k\pi l}{M+1} \frac{1}{1-\mathrm{e}^{-i\frac{2\pi k}{M+1}}}\frac{2}{1+M} g^{-1} \sin \frac{k\pi}{M+1}.
    \label{eq:FT}
\end{equation}
 The expression above can be further simplified by summing first over the index $k$ and leaving the summation over $l$ as last. After some simplifications we get
\begin{equation}
    f(t)=\frac{-1}{2g(M+1)}\sum_{l=1}^M g^{l}[\left(D_M(t+1+l)-D_M(t+l)\right)-\left(D_M(t+1-l)-D_M(t-l)\right)],\label{eq:Otsimple}
\end{equation}
where $D_M(x)=\sum_{k=-M}^{M} \mathrm{e}^{\ii k \frac{\pi x}{M+1}} = \sin((M+\frac{1}{2})\frac{\pi x}{M+1})/\sin(\frac{\pi x}{2M+2})$ is the Dirichlet function \cite{dirichlet}. For any integer $x$, this function simplifies to 
    \begin{align}
	   D_M(x)&=\begin{cases}
	   (-1)^{x+1}, \qquad \mathrm{when} \quad x\neq 2(M+1)p\\
	   2M+1, \qquad \mathrm{when} \quad x=2(M+1)p
	   \end{cases} \\
    &=(-1)^{x+1}+(2M+2)\sum_{p=-\infty}^{\infty}\delta_{x,2(M+1)p},
    \end{align}	
where $p$ is an integer and the remaining sum describes a ``train'' of Kronecker deltas. Substituting the above result in Eq.~\ref{eq:Otsimple} we have
    \begin{equation}
	   f(t)=-\frac{1}{g}\sum_{l=1}^M g^{l}\sum_{p=-\infty}^{\infty}\left[
                   (\delta_{t+1+l,2(M+1)p} -\delta_{t+l,2(M+1)p})
                  -(\delta_{t+1-l,2(M+1)p}-\delta_{t-l,2(M+1)p})\right].
    \end{equation}
From the sum above, we see that as $t$ increases, the behavior of $f(t)$ may change depending upon which of the delta functions click. Let us first assume $0<t<M$, then the only Kronecker deltas that click are for $p=0$. Therefore,
    \begin{align}
        f(t)&=-\frac{1}{g}\sum_{l=1}^M g^{l}\left[(\delta_{t+1+l,0}-\delta_{t+l,0})-(\delta_{t+1-l,0}-\delta_{t-l,0})\right] \nonumber 
        =(g-1)g^{t-1}.
    \end{align}	
For $g>1$ we therefore have the exponential $\sim g^t$ growth, with the rate $g$ being equal to the $\lp$ (Eq.~\ref{eq:lps}). For $M+1<t<2M-1$ the Kronecker deltas that click now are for $p=1$ and therefore, using similar arguments as above, we get an exponential decay \footnote{Note that for $0<g<1$ our choice of $\bra{w}$ and $\ket{v}$ give an initial decay followed by an exponential growth in each period. However, for random $\bra{w}$ and $\ket{v}$ we always observe an initial exponential growth, followed by a decay, regardless of $g$.}
    \begin{align}
        f(t)=(g-1)g^{2(M+1)}g^{-t-2}.
    \end{align}
For times $t>2M$, $f(t)$ is computed by considering its periodicity, $f(t+T)=f(t)$. At $t=M$ and $t=M+1$, that is at the peak when one flips from boom to bust, it is straightforward to see that $f(t)=-g^{M-1}$ and therefore $f(t)$ takes negative values. The sudden jumps to negative values near the peaks are also present for random choices of the vectors $\ket{v}$ and $\bra{w}$.

An alternative way to understand the behavior of $f(t)$ is by taking first the sum over $l$ in Eq.~\ref{eq:FT}, i.e., writing it as a sum over eigenvalues $f(t)=\sum_j c_j \lambda_j^t$, and recognizing that the resulting expression is nothing but a Fourier transformation due to the special form of $\lambda_j$,
\begin{equation}
    f(t) = \sum_{k=-M}^{M} \mathrm{e}^{i \frac{k \pi }{M+1}t} c_k,
\end{equation}
where 
\begin{equation}
    c_k = \frac{i}{g(M+1)} \cdot (-1)^k\frac{ \left(1-e^{\frac{i \pi  k}{M+1}}\right) \left((-1)^k-g^{M+1}\right) \sin \left(\frac{\pi  k}{M+1}\right)}{g+\frac{1}{g}-2 \cos \left(\frac{\pi  k}{M+1}\right)}.
    \label{eq:ck}
\end{equation}
A necessary condition to observe $f(t) \propto \lp^t$ for $t\lesssim M$ is that $c_k$ alternate in sign with $k$ and that  $c_k$ grow exponentially large with the system size \cite{marko23}. In Eq.~\ref{eq:ck}, this can be explicitly seen: for large $M$ and $g>1$ the term $g^{M+1}$ prevails over $(-1)^k$, which makes $c_k$ proportional to $g^M$ and alternate in sign with $k$. These necessary conditions do not hold for $g=1$, where all the coefficients $c_k$ are of the same sign and do not grow exponentially with the system size. The effect therefore comes about due to a delicate cancellation of exponentially large terms in the spectral expansion. Exponentially large terms are in turn possible due to the singular $V$ that diagonalizes $A$. Singular $V$ implies that condition numbers $\kappa_j$ diverge, i.e., the spectrum of $A$ is different than its pseudospectrum.

The Fourier coefficients of $f(t)$, i.e. $c_k$, are fixed by the components of the vectors $\bra{w}$ and $\ket{v}$. Hence, special choices of these vectors can in principle give rise to an arbitrary periodic $f(t)$, one just has to ``choose'' $c_k$ to be equal to the Fourier transformation of $f(t)$. Looking from this angle it might come as a surprise that the observed periodic boom and bust in Fig.~\ref{fig:example} persist also for generic vectors, e.g., for random ones. The resolution is that while the exact calculation performed in this Section is specific to the special choice of vectors, the phenomenon is not. Pseudospectra are robust, i.e., slightly perturbed matrix has the same pseudospectrum, and therefore the norm of $\lVert A^t \rVert$, which is for finite $t$ dominated by the pseudospectrum~\cite{trefethen}, will be observed for generic vectors.

\section{Other examples}
\label{sec:other}

In this Section, we briefly look at other examples of the matrix $A$ which show similar behavior to the one described in the Section \ref{sec:simple}. To observe a periodic boom and bust similar to the one shown in Fig.~\ref{fig:example}, we need to ensure two necessary conditions, namely (i) $A$ must generate periodic behavior, so $A^T=\1$, and (ii) the largest pseudoeigenvalue $\lp$ in the pseudospectrum of $A$ must be greater in modulus than its largest eigenvalue $\l1$. In general, condition (i) is satisfied for some possibly large $T$ by any matrix having eigenvalues on the unit circle at commensurate angles.

Such a spectrum can be also obtained by choosing $A=e^{\ii \alpha H}$, where $H$ is a non-Hermitian matrix with an evenly spaced, say integer spectrum. Non-Hermiticity of $H$ is a necessary condition for (ii) because $\lp>\l1$ cannot happen for a unitary $A$. Such equally spaced so-called picket-fence spectra are sometimes used when testing various spectral statistics, for instance in quantum chaos. In the first subsection, we take $H$ to be the matrix that appears in the description of the Ehrenfest urn (a.k.a. the Ehrenfest model), a simple toy model devised~\cite{ehrenfest} by Paul and Tatiana Ehrenfest to illustrate the 2nd law of thermodynamics. Another example of such $H$ with integer spectrum is considered in Ref.~\onlinecite{bondesson2005matrix}. Behavior similar to the boom and bust cycles can also be observed, up to a scaling factor, in systems where the spectrum does not lie on the unit circle. Such an example, taken from Ref.~\onlinecite{PRX}, is discussed in the second subsection, where we focus on the time evolution of purity in a specific random quantum circuit.

\subsection{Ehrenfest urn matrix}

In this subsection, we study a system where $A=e^{\ii \alpha H}$, where $H$ is a matrix with integer spectrum. Specifically, we take $H$ to be the Ehrenfest urn matrix (also called the Kac-Sylvester matrix), see for instance ~\cite{taussky1991another,edelman1994road,gladwell2014test},

\begin{equation}
	H=\begin{pmatrix}
	0  & 1&0  &0& \dots\dots& 0&0 \\
	N-1 & 0 & 2&0 & \dots\dots&0&0 \\
	0 & N-2& 0 & 3 &\dots \dots &0&0\\
	\vdots&\vdots &\vdots & \ddots&&\vdots&\vdots \\
	\vdots&\vdots&\vdots& & \ddots &\vdots&\vdots \\
	\vdots&\vdots&\vdots&\dots&2&0 & N-1\\
	0&0 & 0&\dots&\dots&1 & 0 
\end{pmatrix}.
\label{eq:H}
\end{equation}
Clearly, $H$ can be interpreted as a hopping matrix of a non-Hermitian tight-binding lattice model in one-dimension, and then $f(t)=\langle w| e^{\ii \alpha H t}| v\rangle$ is the overlap of the quantum state at time $t$ with some state $\ket{w}$. For even (odd) $N$, the eigenvalues of $H$ are odd (even) integers in the interval $[-N, N]$, and therefore the eigenvalues of $A$ lie on the unit circle at angles that are commensurate with $\pi$ for our choice $\alpha=\frac{\pi}{2N}$. The pseudospectrum of $A$ is obtained by exponentiating the pseudospectrum of $\ii \alpha H$, which is given by Eq.~9.2 in Ref.~\onlinecite{trefethen}. The largest values in the pseudospectrum of $\ii \alpha H$ lie on the circle with radius $\pi/2$, hence the largest value in the pseudospectrum of $A$ is equal to $\mathrm{e}^{\pi/2}$.

\begin{figure}[h]
        \begin{center}
        \includegraphics[width=0.7\textwidth,page=6]{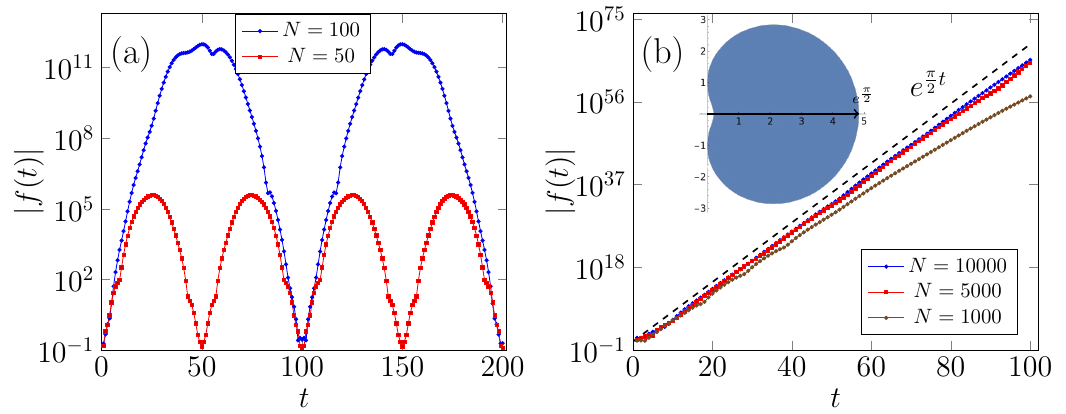} 
        \caption{Dynamics for the Ehrenfest urn matrix from Eq.~\ref{eq:H}. Frame (a) shows periodic exponential growth and exponential decay of $f(t)$. The vectors $\ket{v}$ and $\bra{w}$ are chosen to be normalized random vectors with real components. Frame (b) shows how $f(t)$ initially behaves for large system sizes. The initial behavior of $f(t)$ converges to $\mathrm{e}^{\pi t/2}$ as we increase the system size. Note that $\mathrm{e}^{\pi/2}$ corresponds to the largest value in the pseudospectrum of $A$, shown as a shaded area in the inline plot in frame (b).} 
    \label{fig:earnfest}
    \end{center}
\end{figure}

In Fig.~\ref{fig:earnfest} (a), we show the behavior of $f(t)$ with $t$ for matrix sizes $N=50$ and $N=100$ and $\alpha = \frac{\pi}{2N}$. We observe a periodic behavior of $f(t)$ with period $T=2N$ of alternating growths and decays, similar to the one in Fig.~\ref{fig:example}, although less sharp at similar $N$. Fig.~\ref{fig:earnfest} (b) shows the initial growth of $f(t)$ for $N=1000$, $N=5000$ and $N=10000$. As can be seen, the initial growth of $f(t)$ is exponential and given by $\mathrm{e}^{\pi t/2}$. Note that $\mathrm{e}^{\pi/2}$ is the largest value in the pseudospectrum of $A$, which is shown in the inline plot in Fig.~\ref{fig:earnfest} (b).

\subsection{Purity in random quantum circuits}

\begin{figure}
    \centering
    \includegraphics[width=0.99\textwidth,page=7]{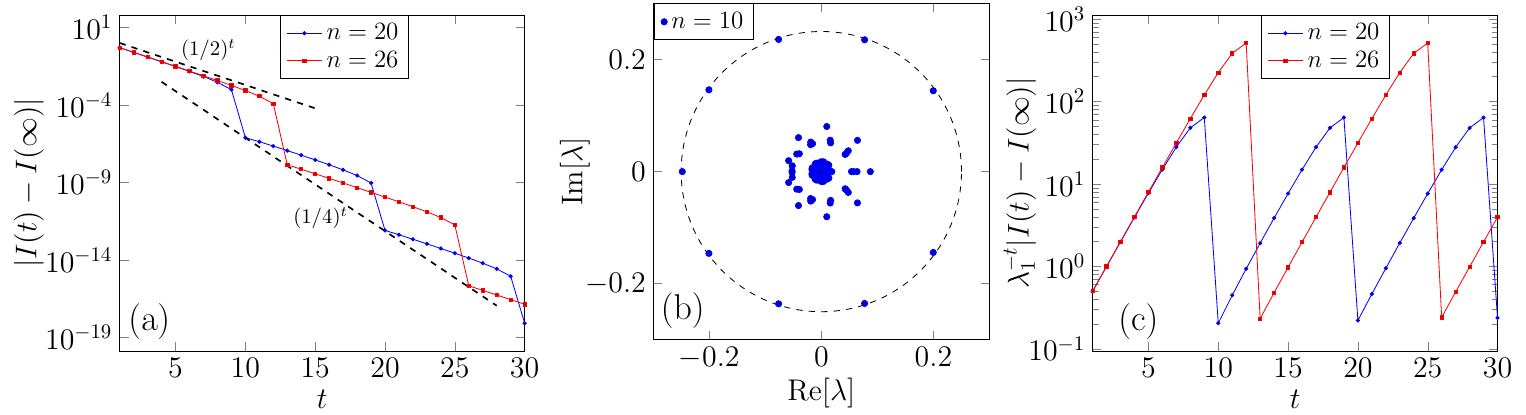}
    \caption{Time evolution of purity $I(t)$ in a staircase open boundary condition random quantum circuit with iSWAP gates. Frame (a) shows how purity of a qubit chain of $26$ and $20$ sites decays exponentially towards its asymptotic value $I(\infty)$. Frame (b) shows the spectrum of the transfer matrix $Z$ used to propagate $I(t)-I(\infty)$ for a chain of $10$ qubits. The dashed circle marks all eigenvalues with modulus $1/4$. Frame (c) shows the propagation of $I(t)-I(\infty)$ by the rescaled transfer matrix $\l1^{-1} \cdot Z$. In this case, purity shows cycles of boom and bust similar to $f(t)$.}
    \label{fig:purity}
\end{figure}

In Ref.~\onlinecite{PRX} an odd behavior of purity has been observed for a particular unitary random circuit in which the gates were arranged in a staircase configuration, with each gate being composed of a fixed two-qubit iSWAP gate and of two random independent single-qubit unitaries. This purity decay is reproduced in Fig.~\ref{fig:purity}(a) (see also Figs. 13(b) and 14(d) in Ref.~\onlinecite{PRX}). The approach of purity $I(t)$ to its stationary long-time value $I(\infty)$ can be written as $I(t)-I(\infty) = \la w | Z^t | v \ra$, see Ref.~\onlinecite{PRX}, and therefore has the same form as $f(t)$ studied here. The spectrum of $Z$, shown in Fig.~\ref{fig:purity}(b), see also Fig.14(a) in Ref.~\onlinecite{PRX}, contains an extensive number of eigenvalues that lie on the circle with radius $\l1=1/4$. These are the largest eigenvalues of $Z$ and are in modulus gapped away from all other (smaller) eigenvalues. In the limit of large time $t$, the behavior of purity should therefore be dominated by the eigenvalues lying on this circle of radius $1/4$ because the contributions from the remaining eigenvalues will be exponentially suppressed in $t$. Thus, the relevant part of the spectrum of the transfer matrix $Z$ is essentially similar to the spectrum of the matrix explored in Section \ref{sec:simple} if we rescale it by a factor of $1/4$. 

Since the transfer matrix is not normal, we know that a naive expectation that the purity will asymptotically decay with the largest eigenvalue, namely $I(t)-I(\infty) \propto (1/4)^t$, is not necessarily correct. This is indeed visible in Fig.~\ref{fig:purity}(a), where the initial decay that lasts up to time proportional to the number of qubits $n$ is $(1/2)^t$. At that point it exhibits an instant jump to a smaller value, and then the story repeats periodically \footnote{The dimension of the transfer matrix $Z$ is $2^n \times 2^n$. This makes the computation of the pseudospectrum of $Z$ difficult. Therefore, it is not confirmed whether the initial decay of purity given by $(1/2)^t$ is connected to the largest value in the pseudospectrum of $Z$.}. The largest eigenvalue $\l1$ still determines the average decay of purity, namely the kinks in the purity decay are located at values $\l1^t$ (similarly, the kinks of $f(t)$ are located at constant values). The influence of $\l1$ can be easily explained by considering the matrix $Z$ rescaled by $\l1^{-1}$. In this case, the ring of largest eigenvalues is located on the unit circle, so $\l1^{-1} \cdot Z$ generates periodic behavior, see Fig.~\ref{fig:purity} (c). Periodicity generated with $\l1^{-1} \cdot Z$ makes $\l1^{-(T+t)} I(T+t) = \l1^{-t} I(t)$, which in turn means that $I(T+t) = \l1^{T} I(t)$.

Purity and $f(t)$ show similar behavior, however, there are differences between these two quantities. Firstly, the period of $f(t)$ is proportional to the matrix size of $A$, while the period of purity is proportional to the number of qubits $n$, even though the matrix size of $Z$ is exponential in $n$. Secondly, $f(t)$ undergoes an exponential growth in the first half of the period, followed by an exponential decay in the second half. Rescaled purity, in contrast, grows exponentially until $T-1$ and then experiences a sudden jump to the initial value. The second part of the periodic behavior of $f(t)$ (as well as the sudden jump of $I(t)$) is still to be understood.

\section{Conclusion}

We found a rather extreme manifestation of pseudospectra in the time dependence of matrix elements $\la w | A^t | v \ra$ for matrices $A$ that are roots of identity, $A^T=\1$, and for which the spectrum and the pseudospectrum differ. While the root condition ensures the exact periodicity in time, and that all eigenvalues lie on the unit circle, the matrix is not normal. We show several instances of such matrices, and in a solvable example demonstrate that half of each period of length $T\approx N$ consists of an exponential growth, immediately followed by an exponential decay. The growth rate is given by the largest value in the pseudospectrum of $A$. If one first takes the limit of matrix size $N$ to infinity, one observes only the growth phase, and therefore dynamics is completely different than no growth one would ``expect'' based on unimodular eigenvalues.

While the exactly solvable case is special, we note that the purity decay with instant jumps observed~\cite{PRX} for a staircase random circuit likely comes from effectively the same phenomenon, even though the matrix there does not have the exactly the same structure studied here. We speculate that the behavior could  be robust and occur in matrices that are not normal and have an extensive number of eigenvalues on a circle, all having diverging condition number, and that are in modulus gapped away from all other eigenvalues (like in Fig.~\ref{fig:purity}(b)). Such a structure of eigenvalues could be a consequence of a symmetry, or could be a result of a degenerate perturbation theory (Puiseux series~\cite{Kato}), like in a breakdown of an extensive Jordan block discussed in Ref.~\onlinecite{marko23}.

\begin{acknowledgements}
We acknowledge support by Grants No.~J1-4385 and No.~P1-0402 from the Slovenian Research Agency.
\end{acknowledgements}

\bibliography{boombust}

\end{document}